\documentclass[aps,nofootinbib,prl,showpacs,class-pre]{revtex4}
\usepackage{setspace} 
\usepackage{amsmath,amssymb,amsfonts,amsthm}
\usepackage{graphicx}
\newcommand{\be}{\begin{equation}}
\newcommand{\ee}{\end{equation}}
\newcommand{\beq}{\begin{eqnarray}}
\newcommand{\eeq}{\end{eqnarray}}

\newcommand{\Mpl}{M_{\rm Pl}}

\newcommand{\dd}{\mathrm{d}}  

\def\lsim{\hbox{ \raise.35ex\rlap{$<$}\lower.6ex\hbox{$\sim$}\ }}
\def\gsim{\hbox{ \raise.35ex\rlap{$>$}\lower.6ex\hbox{$\sim$}\ }} 

\def\xrightarrow#1#2#3#4{\,\lower#1pt\hbox{$\stackrel{\stackrel{\displaystyle
        #2} {\hbox to #3cm{\rightarrowfill}}}{#4}$}\,}

\begin{document}

\begin{flushleft}
KCL-PH-TH/2014-9
\end{flushleft}

\title{How well do we understand the thermal history of the Universe?\\
  Implications of the recent BICEP2 data} \author{Mairi
  Sakellariadou\footnote{mairi.sakellariadou@kcl.ac.uk}}
\affiliation{Department of Physics, King's College London, University
  of London, Strand WC2R 2LS, London, U.K.}

\begin{abstract}
Applying Grand Unified Theories in the context of the early Universe,
cosmic strings are an unavoidable partner of the inflaton field. To
render such models compatible with the CMB temperature anisotropies,
the inflationary scale has to be less than $10^{15}\ {\rm GeV}$, a
value which is inconsistent with the high inflationary scale deduced
from the tensor-to-scalar ratio announced from the BICEP2
collaboration. Some questions may be consequently raised about our
understanding of the thermal history of the Universe.
\end{abstract}


\maketitle

%
Accepting the validity of Grand Unified Theories (GUTs) leads to the
conclusion that the Universe, in the earliest stages of its evolution,
was hotter and consequently in a more symmetric state. Subsequently,
in the expansion process (Hubble's law), its temperature dropped and
the Universe underwent a series of phase transitions followed by
spontaneous breakdown of symmetries. Topological defects could then
be left as remnants. In the context of (local) gauge theories ---
global theories, being rather unphysical, will not be discussed
here --- cosmic
strings~\cite{Vilenkin,Sakellariadou:2006qs,Sakellariadou:2009ev} have
always been considered as the only type of (local) topological defects
compatible with the observed Universe.

To dilute any undesired topological defects --- domain walls,
monopoles or textures --- a stage of cosmological
inflation~\cite{Starobinsky:1980te,Guth:1980zm, Linde:1981mu} is
usually imposed.  Such an inflationary era solves (by construction)
the horizon and flatness problems, while as a bonus, it provides a
mechanism for the origin of adiabatic Gaussian fluctuations, in
consistency with the measured Cosmic Microwave Background (CMB)
temperature anisotropies. Hence, cosmological inflation can account
for the observed large-scale structure, within the framework of the
theory of gravitational instability.

In this spirit, one has to examine defect formation within the series
of phase transitions from a large GUT gauge group G$_{\rm GUT}$ down
to the Standard Model (SM) group ${\rm G}_{\rm SM}={\rm SU(3)}\times
{\rm SU(2)}\times{\rm U(1)}$, with ${\rm G}_{\rm SM}\subset {\rm
G}_{\rm GUT}$, and then postulate an inflationary era at the end of
the (last) formation of any undesired defects.  

In the context of GUTs, it has been commonly suggested the type of
hybrid inflation~\cite{Linde:1993cn} realised within a scheme
\begin{equation}
{\rm G}\stackrel{M_{\rm GUT}}{\hbox to
0.8cm{\rightarrowfill}} {\rm H}_1 \xrightarrow{9}{M_{\rm
infl}}{1}{\Phi_+\Phi_-} {\rm H}_2 \rightarrow {\rm G}_{\rm SM}~,
\end{equation}
where $\Phi_+, \Phi_-$ is a pair of GUT Higgs superfields in
non-trivial complex conjugate representations, which lower the rank of
the group by one unit when acquiring a non-zero vacuum expectation
value. The inflationary stage takes place at the beginning of the
symmetry breaking ${\rm H}_1\stackrel{M_{\rm infl}}{\longrightarrow}
{\rm H}_2$ and is based on a globally supersymmetric renormalisable
superpotential $W$. Note that G stands here for either ${\rm G}_{\rm
  GUT}$ or ${\rm G}_{\rm GUT}\times$U(1).

Performing a rather exhaustive investigation, it has been shown in
Ref.~\cite{Jeannerot:2003qv} that cosmic strings are generically
formed at the end of hybrid inflation.
Hence, provided Grand Unified Theories is a valid hypothesis, cosmic
strings are expected to be left behind, as relics of an earlier more
symmetric phase. Their cosmological consequences --- and therefore
their observational fingerprints --- are mainly based on their
gravitational interactions, which essentially depend on their linear
mass density $\mu\sim T^2$ (with $T$ denoting the
temperature of the phase transition during which strings were formed),
which is intimately related to the scale of inflation.

Switching gears and moving to M-theory, in which branes of various
dimensions are embedded in a higher dimensional space, one can still
build a (brane) inflationary scenario, as the outcome of brane
interactions. Within IIB string theory there are two classes of brane
inflation models: D3/D7 inflation~\cite{Dasgupta:2002ew,
  Dasgupta:2004dw,Gwyn:2010rj}\footnote{It is worth noting the caveat
  discussed in Ref.~\cite{Gwyn:2011tf}.} and brane-antibrane inflation
with D3/${\overline{{\rm D}3}}$~\cite{Kachru:2003sx} being the most
studied example.  Such brane inflation models generically
accommodate~\cite{Sarangi:2002yt,Dvali:2003zj} cosmic
superstrings~\cite{Polchinski:2004hb,Sakellariadou:2008ie}, which will
then play the r\^ole of their field theoretic analogues.

Assuming the validity of the above discussion, one would
expect~\cite{Bouchet:2000hd} that cosmic (super)strings are a
legitimate partner of the inflaton field, both seeding the initial
density fluctuations which lead to the measured CMB temperature
anisotropies. Hence, one may consider~\cite{Bouchet:2000hd}
\begin{equation}
C_\ell =   \alpha C^{\scriptscriptstyle{\rm I}}_\ell
                     + (1-\alpha) C^{\scriptscriptstyle{\rm S}}_\ell~,
\label{cl}
\end{equation}
where $C^{\scriptscriptstyle{\rm I}}_\ell$ and $C^{\scriptscriptstyle {\rm
S}}_\ell$ stand for the (COBE normalised) Legendre coefficients due to
adiabatic inflaton fluctuations and those stemming from a cosmic
(super)string network, respectively; the coefficient $\alpha$ is a free
parameter expressing the relative amplitude for the two contributions.

Such theoretical constructions have to be tested against observational
data and astrophysical measurements, and this has been the aim of
various studies, in particular with respect to the CMB
data~\cite{Ade:2013kta,Ade:2013xla}. Let us leave aside the precise
predictions of a given model, and summarise the generic predictions of
each class of models (strings, and inflation), which turn out to be
intrinsically different for reasons which are known for about two decades.

Cosmic (super)string models lead to isocurvature density perturbations
--- the total density perturbation vanishes, whilst the density
perturbation of individual particle species does not. More
importantly, perturbations are generated continuously; they evolve
according to inhomogeneous linear perturbation equations, with the
energy momentum tensor of strings being determined by their non-linear
evolution. The resulting angular power spectrum of temperature
anisotropies is qualitatively different than the one obtained via
adiabatic perturbations from an inflationary model. More precisely,
one obtains a broad low peak instead of the distinct series of
acoustic peaks. Moreover, string models predict generically
non-Gaussianities in the CMB temperature spectrum.

The first year Planck data have been analysed~\cite{Ade:2013xla}
within the framework of $\Lambda$CDM cosmology for a simple adiabatic
model with an extra string contribution expressed in terms of the
fractional contribution, $f_{10}$, of strings to the CMB temperature
spectrum at multipole $\ell=10$. For an Abelian-Higgs (AH) field theory
model, the constraint turned out to be $f_{10}<0.028$, or equivalently
$G\mu_{\rm AH}/c^2<3.2\times 10^{-7}$~\cite{Ade:2013xla}.  Moreover,
the Planck data impose severe constraints on any primordial
non-Gaussianity~\cite{Ade:2013ydc}.
The evident question is whether the theoretical models can be
compatible with the observational data~\cite{Sakellariadou:2013wwa}.

The scalar potential in supergravity has the general form~\cite{nilles}
\begin{equation}\label{DpotenSUGRA}
V=\frac{e^G}{M_{\rm Pl}^4}\left[G_i(G^{-1})^i_jG^j-3 \right] + \frac{1}{2} 
[{\rm Re} f(\Phi_i)]^{-1} \sum_a g_a^2 D_a^2~,
\end{equation}
where 
\begin{equation}
G=\frac{K}{M_{\rm Pl}^2}+\ln\frac{|W|^2}{M_{\rm Pl}^6}~,
\end{equation}
with the K$\ddot{\rm a}$hler potential $K(\phi, \phi^*)$ being a real
function of the scalar components of the chiral superfields $\Phi_i$
and their Hermitian conjugates.  Upper (lower) indices $(i,j)$ denote
derivatives with respect to $\phi_i$ (${\phi^i}^*$), namely
\begin{equation}
G^i\equiv \frac{\partial G}{\partial \phi_i}~, 
\quad G_j\equiv \frac{\partial G}{\partial {\phi^j}^*}~.
\end{equation}
$f(\Phi_i)$ is the gauge kinetic function, and $g_a$ is the coupling
of the $U(1)^a$ symmetry, which is generated by ${T_a}$ and under
which the chiral superfields $S$ (with $S$ a gauge singlet playing
the r\^ole of the inflaton), $\Phi_+$ and $\Phi_-$ have charges $0$,
$+1$ and $-1$ respectively. Finally,
\begin{equation}
D_a=\phi_i {(T_a)^i}_j K^j+\xi_a~,
\end{equation}
where $\xi_a$ is the Fayet-Iliopoulos term; it can only be nonzero if
$T_a$ generates a U(1) group.  

Considering one-loop radiative corrections to the scalar potential
along the inflationary valley, the effective scalar
potential~\cite{DvaShaScha,Rocher:2004et} in the context of minimal
supersymmetric hybrid inflation with minimal K\"ahler potential
\begin{equation}
K=|s|^2+|\phi_+|^2 +|\phi_-|^2~,
\end{equation}
lead to a large spectral index which is not preferred from the recent
CMB measurements. Note that $s, \phi_1, \phi_2$ are the bosonic
components of the superfields $S, \Phi_1,\Phi_2$. 

However, as it has been shown in Ref.~\cite{BasteroGil:2006cm},
supersymmetric hybrid inflation with non-minimal K\"ahler potential,
including radiative corrections obtained through the one-loop
effective potential, may lead to a red-tilted spectrum ($n_{\rm
  s}\approx 0.96$), in agreement with the WMAP~\cite{Hinshaw:2012aka}
and Planck~\cite{Planck:2013nga} data. In particular, it was
shown~\cite{BasteroGil:2006cm} that considering a non-minimal K\"ahler
potential
\begin{equation}
K =|s|^2+|\phi_+|^2 +|\phi_-|^2+\kappa_s{|s|^4\over
M_{\rm Pl} ^2}+\kappa_{s\phi_+}{|s|^2|\phi_+^2|\over M_{\rm
Pl}^2} +\kappa_{s\phi_-}{|s|^2|\phi_-|^2\over M_{\rm
Pl}^2}+\cdots~,
\label{nonminV}\end{equation}
one can get a red-tilted spectrum.  The consistency of this model
(based on the K\"ahler potential Eq.~(\ref{nonminV})) with the CMB
imposed constraints on the cosmic string contribution was first
performed in Refs.~\cite{Rocher:2006nh,Rocher:2004my}.

For the non-minimal K\"ahler potential above, Eq.~(\ref{nonminV}), the
effective potential reads~\cite{Rocher:2006nh}
\begin{equation}
V_{\rm eff}(|s|)=\frac{g^2\xi^2}{2}\left\{ 1+\frac{g^2}{16\pi^2}\left[
  2\ln \left( z\frac{g^2\xi}{\Lambda^2}\right)+f_V(z) \right] \right\}
~,
\end{equation}
where 
\begin{equation}\label{deffV}
f_V(z) \equiv (z+1)^2\ln\left( 1+\frac{1}{z}\right) + (z-1)^2\ln\left(
1-\frac{1}{z}\right)~
\end{equation}
with
\begin{equation}
z\equiv \frac{\lambda^2|s|^2}{g^2\xi}\exp\bigg(\frac{|s|^2}{\Mpl^2}
+\kappa_{\rm s}\frac{|s|^4}{\Mpl^4}\bigg) \frac{1}{(1+f_+)(1+f_-)}~,
\end{equation}
and 
\begin{equation}
f_+\equiv\kappa_{s\phi_+}{|s|^2\over M_{\rm
Pl}^2}\ \ ; \ \ f_-\equiv\kappa_{s\phi_-}{|s|^2\over M_{\rm
Pl}^2}~.
\end{equation}
The number of e-folds is calculated from
\begin{equation}\label{efoldings}
N_{\rm Q}\equiv \ln \left( \frac{a_{\rm end}}{a_Q}\right) =
\frac{8\pi^2}{g^2 M_{\rm Pl}^2}\int_1^{z_Q} \frac{\dd z}{z^2
f^2_z[s(z)] f_{V'}(z)}~,
\end{equation}
with the index $_{\rm Q}$ denoting the scale
of the CMB quadrupole anisotropy, and the definitions
\begin{eqnarray}\label{deffV'}
f_{V'}(z)&\equiv& (z+1)\ln\left( 1+\frac{1}{z}\right) +
(z-1)\ln\left( 1-\frac{1}{z}\right) ~,
\\
f_z(|s|)&\equiv&2|S|\left[ \frac{1}{\Mpl^2}+ \frac{2\kappa_{\rm s}|s|^2}{\Mpl^4}
  +\frac{1}{|s|^2}-\frac{\kappa_{s\phi_+}}{(1+f_+)\Mpl^2}-\frac{\kappa_{s\phi_-}}{(1+f_-)\Mpl^2}\right]~.
\label{newfz}
\end{eqnarray}
Setting the number of e-folds to about 60, one can fix the value of
the inflaton field $s_{\rm Q}$.  The scalar and tensor part
contributions of the inflaton field to the temperature anisotropy
read~\cite{Rocher:2006nh}
\begin{eqnarray}
\left(\frac{\delta T}{T}\right)_{\mathrm{Q-scal}} &\simeq&
\frac{1}{4\sqrt{45} \pi} \frac{V^{3/2}(s_{\rm Q})}{M_{\rm
Pl}^3V'(s_{\rm Q})} \nonumber\\
&\simeq&
\frac{\sqrt{2}\pi}{\sqrt{45}}\frac{\xi}{g}\, \frac{1}{M_{\rm
Pl}^3}\, z_{\rm Q}^{-1}f_{V'}^{-1}(z_{\rm Q})f_z^{-1}(s_{\rm Q})~.
\label{dTsurT}\end{eqnarray}
and
\begin{eqnarray}
\left(\frac{\delta T}{T}\right)_{\mathrm{Q-tens}} &\simeq&
\frac{(0.77)}{(8\pi)} \frac{V^{1/2}(s_{\rm Q})}{M_{\rm Pl}^2}\nonumber\\
&\simeq&
\frac{0.77}{8\sqrt{2}\pi}\frac{1}{M_{\rm Pl}^2} g \xi~,
\label{inflsctens}
\end{eqnarray}
respectively.
The tensor over scalar ratio, $r_{\rm infl}$, is given by
\begin{equation}
r_{\rm infl} = \frac{0.77\sqrt{45}}{16\pi^2}\, g^2\, z_{\rm Q} \,
M_{\rm Pl}\, f_{V'}(z_{\rm Q}) f_z(s_{\rm Q})~.
\end{equation}
One can then compute the cosmic string contribution to the quadrupole
CMB temperature anisotropy, as a function of the superpotential
coupling $\lambda$, for various values of $g$ and $\kappa_{s,
  \phi_\pm}$, which are considered as parameters. The allowed
parameter space can be expressed as a constraint on the
Fayet-Iliopoulos parameter $\xi$, noting that the quadrupole
contribution to the CMB temperature anisotropies from the cosmic
strings formed at the end of hybrid inflation is
\begin{equation}
\left(\frac{\delta T}{T}\right)_{\mathrm{Q-CS}} \simeq
\frac{9}{4}\xi ~.
\end{equation}
Performing a similar analysis as in Ref.~\cite{ Rocher:2004my} we obtain
\begin{equation}\label{limit-xi}
\sqrt\xi\lsim 10^{15}\ {\rm GeV}~.
\end{equation}
This limit is inconsistent with the BICEP2 results~\cite{Ade:2014xna} on the inflationary
scale. Let us emphasise that computing the mass scale of symmetry
breaking, given by $\sqrt\xi$, one finds that it increases with
$\lambda$. Hence, to achieve compatibility with the inflationary scale
set by BICEP2, one needs to increase the value of $\lambda$ above the
upper limit (a few $\times 10^{-5}$) imposed from the CMB temperature
anisotropies measurements. 

Summarising, considering hybrid inflation with non-minimal K\"ahler
potential one is able --- through the new degrees of freedom, namely
the parameters $\kappa_{\rm s}$ and $\kappa_{\phi_\pm}$ --- to
accommodate the Planck and BICEP2 data on the spectral index and the
tensor-to-scalar ratio, respectively. However, the new parameters do
not improve the constraint on the allowed string energy scale, which
leads to a low inflationary scale.  Note that the constraints will not
improve by considering hybrid inflation leading to semi-local
strings~\cite{Urrestilla:2004eh}, since similar CMB
constraints~\cite{Urrestilla:2007sf} are also imposed to such a string
network.  Hence, there is an inconsistency between mixed ``hybrid
inflation + string'' models and current experimental data. To render
the models compatible with the constraints one should lower the string
energy scale, a trick that is no longer possible given the GUT-scale
inflationary energy implied from the BICEP2 measurements, provided
these measurements are indeed
confirmed~\cite{Mortonson:2014bja,Flauger:2014qra}.  More precisely,
the BICEP2 analysis was based on dust polarization models that
predicted subdominant contamination of their B-mode signal by dust
polarization. Very recently, the Planck collaboration, using the
Planck HFI polarization data, concluded~\cite{Adam:2014bub} that there
is a larger uncertainty in the B-modes produced by dust, implying the
need for assessment of the polarized dust signal even in the cleanest
windows of the sky. As a result, BICEP2 must be treated as an upper
bound on primordial B-modes rather than a detection.

Turning to the brane inflationary models, one expects similar
constraints on the cosmic superstring contribution to the CMB
temperature anisotropies. This is due to the expected broad low peak,
a generic characteristic of all models with {\sl seeds}, defined as
any non-uniformly distributed form of energy, which contributes only a
small fraction to the total energy density of the Universe and which
interacts with the cosmic fluid only gravitationally.

In short, combining the temperature anisotropies data with the results
on the CMB polarisation B-modes, we unavoidably conclude that mixed
``hybrid inflation + string'' models face severe problems. But such
models have been suggested as a realistic outcome of Grand Unified
Theories applied in the context of the early Universe. Moreover, we
have recently examined~\cite{Cacciapaglia:2013tga} whether
supersymmetric hybrid inflation can be (naturally) embedded within the
minimal SO(10) model. We have shown~\cite{Cacciapaglia:2013tga} that
none of the singlets of the Standard Model symmetries in the minimal
set of SO(10) representations, can satisfy the necessary conditions
for a scalar field to play the r\^ole of the inflaton; a result that
probably remains valid for other gauge groups beyond SO(10).

It is therefore clear that we are missing some important elements in
order to reassemble the puzzle of the early Universe evolution.  It
may be worth analysing the CMB data beyond the $\Lambda$CDM model, or
further investigating inflationary models with a few e-folds after
string formation~\cite{Kamada:2012ag}.


\end{document}